\documentclass[12pt]{iopart}

\usepackage{graphicx}
\usepackage{amstext, amssymb}
\usepackage{braket}
\usepackage{bbm}
\usepackage{bbold}
\usepackage{wasysym}
\usepackage{cleveref}

\newcommand{\id}{\mathbb{1}}    
\newcommand{\calD}{\mathcal{D}} 
\newcommand{\calS}{\mathcal{S}} 
\newcommand{\calL}{\mathcal{L}} 
\newcommand{\calP}{\mathcal{P}} 
\newcommand{\eff}{\text{eff}}   

\begin{document}

\title{Asymptotic Floquet states of open quantum systems: The role of interactions}

\author{M Hartmann$^1$, D Poletti$^2$, M Ivanchenko$^3$, S Denisov$^{1,3,4}$ and P H\"anggi$^{1,3,4,5,6}$}
\address{$^{1}$Institute of Physics, University of Augsburg, Universit\"{a}tsstra{\ss}e 1, 86159 Augsburg, Germany}
\address{$^{2}$Singapore University of Technology and Design, 8 Somapah Road, 487372 Singapore}
\address{$^{3}$Department of Applied Mathematics, Lobachevsky State University of Nizhny Novgorod, Gagarina Av. 23, Nizhny Novgorod, 603950, Russia}
\address{$^{4}$Nanosystems Initiative Munich, Schellingstra{\ss}e 4, 80799 M\"unchen, Germany}
\address{$^{5}$Department of Physics, National University of Singapore, Singapore 117546}
\address{$^{6}$Centre for Quantum Technologies, National University of Singapore, 3 Science Drive 2, Singapore 117543}

\begin{abstract}
When a periodically modulated many-body quantum system is weakly coupled to an
environment, the combined action of these temporal modulations and dissipation
steers at long times the system towards a state characterized by a
time-periodic density operator. To resolve this asymptotic non-equilibrium
state at stroboscopic instants of time, we introduce the dissipative Floquet
map and evaluate the stroboscopic density operator as its invariant. Particle
interactions control properties of the map and thus the features of its
invariant. In addition, the spectrum of the map provides insight into the
system relaxation towards the asymptotic state and may help to understand
whether it is possible (or not) to construct a stroboscopic time-independent
Lindblad generator which mimics the action of the original time-dependent one.
We illustrate the idea with a scalable many-body model, a periodically
modulated Bose-Hubbard dimer. We contrast the relations between the
interaction-induced bifurcations in a mean-field description with the numerical
exact stroboscopic evolution and discuss the characteristics of the genuine
quantum many-body state {\it vs} the evolution of its mean-field counterpart.
\end{abstract}

\pacs{03.75.Gg, 42.50.Dv, 05.45.Mt}

\noindent{\textit{Keywords} open quantum system, asymptotic state, many-body, Floquet states}

\submitto{\NJP}
\maketitle

\section{Introduction}

Many-body effects in combination with a coupling to an environment give rise to
a variety of phenomena which are of beneficial use for quantum technologies.
Interactions sculpt the spectrum of different collective states and moderate
transitions between them \cite{sachdev}. Effects of the system-environment
coupling, however weak they are, play a decisive role in out-shaping the
system's asymptotic state. Indeed, such effects may not necessarily present a
nuisance but can be as well of practical use. Particularly, they can be
exploited to steer the system towards desired states, including pure and
high-entangled ones \cite{wolf, DiehlZoller2008}. This recent idea of
engineering by dissipation \cite{KrausZoller, Bardyn, barr, kienz, ketz} has
promoted a dissipative time evolution of the system dynamics to the same level
of importance as that obtained with a unitary evolution.

Time periodic modulations can also strongly modify the state of a quantum
system. In the coherent limit, time-periodic modulations implicate an explicit
time-periodicity of the Hamiltonian, i.e., $H(t+T) = H(t+2\pi/\omega) = H(t)$,
where $T$ is the driving period and $\omega$ is the frequency of modulations.
The system dynamics is governed by the Floquet states \cite{shirley,sambe,
grifoni}; i.e., the eigenstates of the unitary Floquet propagator $U_T = {\cal
T} \exp\left[-\frac{\rmi}{\hbar} \int_0^T H(\tau) \mathrm{d}\tau\right]$, where
$\cal T$ is the time-ordering operator. The particular structure of the Floquet
propagator, and thus the properties of the Floquet states, depend on modulation
parameters. This allows to grasp effects \cite{grifoni, kohler, arimondo,
bukov, top, eckardt, santos, nagerl, TsujiAoki2008, eck1} which are out of
reach of experimentally available time-independent Hamiltonians in atom optics,
optomechanics, and solid state physics.

Our key objective here is to investigate the combined effect of all three
factors, namely (i) many-body interactions, (ii) coupling to an environment,
and (iii) periodically varying external driving. We start by introducing the
notion of Floquet maps which constitute an extension to the dissipative case
\cite{grifoni} of the unitary Floquet propagator \cite{shirley,sambe}, and
demonstrate how those can be used to obtain non-equilibrium asymptotic states.
We next study a many-body model, a driven Bose-Hubbard dimer, and use both a
full quantum mechanical treatment and a mean-field description, to gain insight
into the properties of the time-periodic asymptotic states.

\section{Dissipative Floquet maps}

We consider the dynamics of a general $M$-dimensional system modeled with a
quantum master equation whose generator $\calL$ is of Lindblad form
\cite{lind,gorini, alicki,book}
\begin{equation}
\label{lind}
\frac{\mathrm{d}}{\mathrm{d}t} \varrho = \calL_t(\varrho) = -\rmi \left[H(t),\varrho\right] + \calD_t(\varrho).
\end{equation}
The first term on the r.h.s.\ describes the unitary evolution of the system's
density operator $\varrho$, governed by the time-periodic Hamiltonian $H(t)$.
The dissipator
\begin{equation}
\label{dissipator}
\calD_t(\varrho) = \sum_{l,k=1}^{M^2-1} \gamma_{kl}(t) \left[V_k\varrho V^\dagger_l - \frac{1}{2}\{V^\dagger_lV_k,\varrho\}\right]
\end{equation}
is built from the set of operators $\{V_k\}$, which, together with the
normalized identity, $V_0 = \id/M$, span the Hilbert-Schmidt space
$\mathcal{B}_{M}$ of the operators acting in $M$-dimensional Hilbert space
\cite{kraus}. Note that all parameters of the system are scaled with respect to
the Planck constant $\hbar$.

Strictly speaking, modulations affect both the unitary and the dissipative part
of the generator $\calL_t$, and in general both the Hamiltonian $H(t)$ and the
dissipative matrix $\Gamma(t) = {\gamma_{kl}(t)}$, become time-dependent
\cite{grifoni,kohler,alicki,pre1997}. The complete theoretical foundation of a
time-dependent master equation of the type (\ref{lind}) still remains an open
problem \cite{spohn1,lendi}. However, if the matrix $\Gamma (t)$ is positive
semi-definite at any instant of time, the propagator $\calP_{s,t} = {\cal T}
\exp(\int_s^t \calL_\tau \rmd\tau)$ is completely positive and
trace-preserving. In this case a master equation in the form of (\ref{lind}) is
meaningful \cite{alicki}. This in turn provides a set-up which is frequently
employed to model quantum systems operating far from equilibrium; see, e.g.,
\cite{ketz,katz,prosen,Shnirman}.

When the generator $\calL_t$ is time dependent, the propagator $\calP_{s,t}$
depends on both the starting time $s$ and the final time $t$.  The closure of
the set of propagators for different times is lost and they no longer form a
semi-group. It is stated by Lendi ~\cite{lendi} that ``the best chance to find
a solution to a master equation [with a time-dependent generator] is only
offered by a possible existence of transformations which eliminate the time
dependence''. Consistently, most studies until now have focused on removing the
time-dependence when dealing with time-periodic generators, either by (i)
finding a proper gauge, which makes the original time-periodic Hamiltonian
time-independent \cite{alicki,chan} and then assuming that the dissipator
remains time-independent in the new frame (this is often a good approximation
in quantum optics, where frequencies of modulations are much higher than the
decay rates), or (ii) by changing to the Floquet basis of the Hamiltonian
$H(t)$ and then performing an additional secular approximation \cite{ketz}, or
(iii) by constructing Magnus expansion-like approximations \cite{mag}. All
these strategies result in deriving an effective time-independent generator
$\calL_\eff$ of the Lindblad form. Once the time-dependence is removed, one has
to calculate the kernel of $\calL_\eff$ to find the asymptotic state
$\varrho_{\infty}$ of the system, i.e., $\calL_{\eff}\varrho_{\infty} =
\mathbb{0}$. Under fairly general conditions \cite{alicki,spohn}, the
time-homogeneous propagator $\calP_t = \exp(\calL_{\eff}t)$ relaxes towards a
unique attractor $\varrho_{\infty}$ of the dissipative quantum evolution.
However, the above discussed approximations cannot always be justified away
from the case of high frequency driving. Below we propose an approach which
does not demand the reduction to such a time-independent form and thus avoids
those corresponding approximation schemes. Note also that it is not necessary
to switch to the Floquet basis of the driven system Hamiltonian.

Because the master equation (\ref{lind}) is manifest linear, we can in the case
of a time-periodic generator $\calL_t$ readily resort to the Floquet theorem
\cite{floquet,floquet1}. We concentrate next on the propagator
$\calP_{\mathrm{F}} \equiv \calP_{0,T} = {\cal T} \exp(\int_0^T \calL_\tau
\mathrm{d}\tau)$ which we refer to as the \textit{Floquet map}. To the best of
our knowledge, the notion of the dissipative Floquet map has not been
introduced in the context of the Lindblad formalism before. The Floquet map
possesses at least one (possibly degenerate) eigenvalue $1$ and all other
eigenvalues lie inside the unit circle. Assuming that the Floquet map is
irreducible \cite{wolf}, the attractor $\varrho_1$ of the map is given by its
fixed point, i.e., the eigen-operator corresponding to the eigenvalue $1$,
$\calP_{\mathrm{F}}\varrho_{1} = \varrho_{1}$. More generally, the number of
different attractor solutions is directly related to the symmetries of the
generator $\calL_t$ \cite{unique}. Particularly, in absence of such additional
symmetries the resulting asymptotic attractor assumes the unique fixed point
solution. Therefore, after a sufficiently large time span any initial density
operator $\varrho_0(t_0)$ will converge to the time-periodic asymptotic state
$\varrho_\mathrm{a}(t)$, i.e., $\varrho_0(t_0) \to \varrho_\mathrm{a}(mT+t_0)$
for an integer $m \gg 1$. The operator $\varrho_1 = \varrho_\mathrm{a}(0) =
\varrho_\mathrm{a}(T)$ is the asymptotic density operator of the system at the
stroboscopic instants of time. Because $\varrho_\mathrm{a}(t)$ is periodic in
time, $\varrho_{\mathrm{a}}(mT + s) = \varrho_{\mathrm{a}}(s)$, the asymptotic
density matrix for any instance of time $t$ can be calculated via propagating
$\varrho_{\mathrm{a}}(0)$ up to time $s$.

\section{Model study: Bose-Hubbard dimer}

To exemplify the ample physics expected to emerge from the interplay of
many-body interaction, dissipation and periodic driving, we consider a system
composed of $N$ interacting bosonic atoms hopping over a dimer which is
subjected to periodic driving. We consider the system Hamiltonian
\begin{equation}
\label{eq:Hamiltonian}
H(t)= -J \left( b_1^{\dagger} b_2 + b_2^{\dagger} b_1 \right) + \frac{U}{2} \sum_{j=1,2} n_j\left(n_j-1\right) + \varepsilon(t)\left(n_2 - n_1\right)
\end{equation}
where $J$ denotes the tunneling amplitude, $U$ is the interaction strength, and
$\varepsilon(t)$ presents the modulation of the local potential. In particular
we choose $\varepsilon(t)=\varepsilon(t+T)=\mu_0+\mu_1\sin(\omega t)$, where
$\mu_0$ presents a static and $\mu_1$ models a dynamic energy offset between
the two sites. Here, $b_j$ and $b_j^{\dagger}$ are the annihilation and
creation operators of an atom at site $j$, and $n_j=b_j^{\dagger}b^{}_j$. This
Hamiltonian has been previously studied theoretically in \cite{coherent, Vardi,
Witthaut, PolettiKollath2012} and has been implemented in several recent
experimental studies \cite{oberthaler, ober1}. However, to the best of our
knowledge, the joint action of all three ingredients -- interaction,
dissipation and temporal driving -- has not been addressed before.

With the coupling constant $\gamma$ taken to be time-independent we use with
$\calD_t = \calD $ the single jump operator \cite{DiehlZoller2008, zoller}
\begin{equation}
\label{eq:jump}
V=(b_1^{\dagger} + b_2^{\dagger})(b_1-b_2) \,.
\end{equation}
This dissipator tends to ``synchronize'' the dynamics on the dimer sites by
constantly recycling anti-symmetric out-phase modes into the symmetric in-phase
ones. Note that our particular setup serves as an illustration only. The
Floquet map approach applies equally well to other cases, e.g., when both parts
of the generator, i.e., the unitary and dissipative parts both are
time-periodic or when there are several jump operators acting on the system.
Because the jump operator (\ref{eq:jump}) is non-Hermitian, the propagators
$\calP_{s,t}$ are not unital and the attractor solution is not the maximally
mixed state, $\varrho_\mathrm{a} \ne \id/M$.

\begin{figure}
\begin{center}
\includegraphics[width=0.99\textwidth]{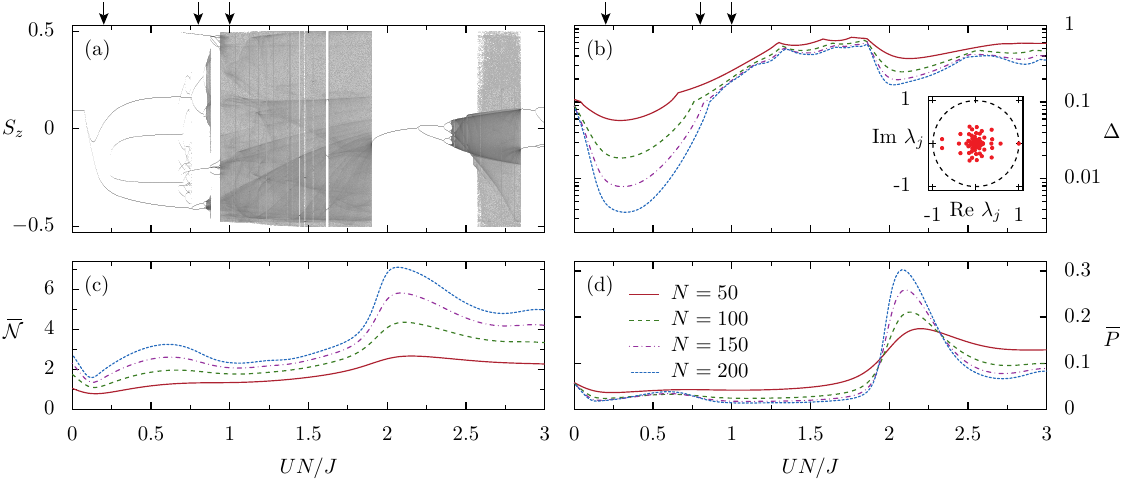}
\end{center}
\caption{(a) Bifurcation diagram for the stroboscopic mean field values of
$S_z=\braket{\calS_z}$ as a function of the interaction strength of the
mean-field equations
(\ref{eq:mean_Sx}--\ref{eq:mean_Sz},\ref{eq:mf1},\ref{eq:mf2}). The arrows
indicate the three regimes presented in figure \ref{fig:poincare}. (b) Spectral
gap $\Delta$, (c) time averaged negativity $\bar{\mathcal{N}}$, and (d)
time-averaged purity $\bar{P}$ of the dissipative Floquet map versus the
interaction strength for different particle numbers $N$. The time-averaged
negativity $\bar{\mathcal{N}}$, the time-averaged purity $\bar{P}$, and the
spectral gap $\Delta$ are defined by
(\ref{eq:purity_avg},\ref{eq:neg_avg},\ref{eq:gap}). The inset in panel (b)
depicts the eigenvalues $\lambda_j$ of the Floquet map $\calP_{\mathrm{F}}$ for
$N=100$ and $UN/J=1$. The other parameters are $\mu_0/J=1$, $\mu_1/J=3.4$,
$\omega/J=1$, $\gamma N/J=0.1$.} \label{fig:bifurcation}
\end{figure}

To gain additional insight into the physics of the model, we derive a set of
mean-field equations and compare its attractor solutions with those of the
quantum Floquet map $\calP_{\mathrm{F}}$. For the dimer problem, it is
convenient to recast the master equation (\ref{lind}) in terms of the spin
operators
\begin{equation}
\label{eq:SxSySz}
\fl
\qquad
\calS_x=\frac{1}{2N}\left(b^{\dagger}_1 b_2 + b^{\dagger}_2 b_1\right), \quad
\calS_y= -\frac{\rmi}{2N}\left(b^{\dagger}_1 b_2 - b^{\dagger}_2 b_1\right), \quad
\calS_z=\frac{1}{2N}\left(n_1 - n_2\right),
\end{equation}
and then study their evolution in the Heisenberg picture \cite{book}. For a
large number of atoms $N \gg 1$, the commutator
$\left[\calS_x,\calS_y\right]=\rmi {\calS_z}/{N}$ becomes negligible small and
similarly for other cyclic permutations. Replacing operators with their
expectation values, $\braket{\calS_k} =\tr [\varrho \calS_k]$, and denoting
$\braket{\calS_k}$ by $S_k$, we end up with
\numparts
\begin{eqnarray}
\label{eq:mean_Sx}
\frac{\rmd S_x}{\rmd t} &= 2\varepsilon(t)S_y - 2UN S_zS_y + 8\gamma N\left(S_y^2+S_z^2\right), \\
\label{eq:mean_Sy}
\frac{\rmd S_y}{\rmd t} &= -2\varepsilon(t)S_x + 2UN S_xS_z +2JS_z - 8\gamma N S_xS_y, \\
\label{eq:mean_Sz}
\frac{\rmd S_z}{\rmd t} &= -2JS_y - 8\gamma N S_xS_z,
\end{eqnarray}
\endnumparts
where we have neglected terms proportional to $\gamma$ of lower order in $N$.
The replacement of operators by their expectation values is justified provided
that $\braket{AB}_t \approx \braket{A}_t\braket{B}_t $. This is not guaranteed
{\it a priori}, and, for a dissipative system, the commutator behaves
differently compared to the unitary setup. A necessary favorable comparison
with the results of the exact quantum analysis then justifies the validity of
this mean field approximation.

The structure of the mean field equations in
(\ref{eq:mean_Sx}--\ref{eq:mean_Sz}) implies that
$\frac{\mathrm{d}}{\mathrm{d}t}S^2=0$. Therefore the quantity
$S^2=S_x^2+S_y^2+S_z^2$ is a constant of motion. This is consistent with the
preservation of the total number of bosons $N$; cf.\ the definitions given by
(\ref{eq:SxSySz}). We therefore can reduce the mean-field evolutions to the
surface of a Bloch sphere; i.e.,
\begin{equation}
\label{eq:mf1}
\left(S_x,S_y,S_z\right) = \frac{1}{2} \left[ \cos(\varphi)\sin(\vartheta),\sin(\varphi)\sin(\vartheta), \cos(\vartheta) \right],
\end{equation}
yielding the equations of motion
\begin{eqnarray}
\label{eq:mf2}
\frac{\mathrm{d}}{\mathrm{d}t} \vartheta &= 2J\sin(\varphi) + 4\gamma N \cos(\varphi)\cos(\vartheta), \nonumber\\
\label{eq:thetaphidot}
\frac{\mathrm{d}}{\mathrm{d}t} \varphi &= 2J\frac{\cos(\vartheta)}{\sin(\vartheta)}\cos(\varphi) - 2\varepsilon(t) +UN \cos(\vartheta) - 4\gamma N \frac{\sin(\varphi)}{\sin(\vartheta)} \,.
\end{eqnarray}

We next analyze the quantum dynamics by using both the Floquet map computed via
(\ref{lind}--\ref{eq:jump}) and contrast the results with the mean-field
equations (\ref{eq:thetaphidot}).

To construct the Floquet map, we use the standard scheme of a vectorization of
the density matrix, which allows to transform (\ref{lind}) into a system of
linear differential equations with time-periodic coefficients. The Floquet map
is obtained by propagating the $\delta$-Kronecker basis \cite{wolf1} over the
full period $T$. Finally, the asymptotic density matrix is given as the
eigen-element of the map corresponding to the unique eigenvalue one.

To extract the classical attractor solution of the mean-field system, we evolve
(\ref{eq:thetaphidot}) from randomly chosen initial conditions, and, after a
transient time $10^4T$, record the value of $S_z$ at the next $250$
stroboscopic instants of time. The so obtained bifurcation diagram is presented
in figure \ref{fig:bifurcation}(a). As the interaction $UN$ varies, we detect
regions containing limit cycles of different periods, chaotic attractors, and
transitions between them \cite{Ott}. We foresight that different dynamical
regimes of the mean-field description are characterized by significantly
different properties of the system in the quantum limit for $N \gg 1$. To check
this hypothesis, we calculate the time-averaged purity
\begin{equation}
\label{eq:purity_avg}
\bar{P} =\frac{1}{T} \int_0^{T} \tr [\varrho_a(t)^2] \mathrm{d}t,
\end{equation}
and also the time-averaged negativity
\begin{equation}
\label{eq:neg_avg}
\bar{\mathcal{N}} = \frac{1}{T} \int_0^{T} \mathcal{N}[\varrho_a(t)] \mathrm{d}t \,.
\end{equation}
Here, $\mathcal{N}$ represents the negativity defined by \cite{Marzolino}
\begin{equation}
{\cal N}[\varrho] = \frac{1}{2} \sum_{k \ne l} |\varrho_{k,l}|
\end{equation}
which characterizes the degree of entanglement in a two-mode system of $N$
indistinguishable bosons. Figures \ref{fig:bifurcation}(c,d) show the
dependence of the two quantities as functions of the interaction strength. It
is interesting that, as the number of bosons increases, changes of the
time-averaged purity and negativity become more pronounced in the vicinity of
bifurcations of the mean-field equations.

The inverse tangent bifurcation \cite{Ott} near $UN/J = 2$ (the transition from
chaos to a period-one limit cycle) is striking: Both the negativity and the
purity of the asymptotic state move to higher values at this point. This
relates to the concept of 'dissipative engineering' used to shape a stationary
many-body system into a pure highly-entangled equilibrium state with the help
of specially designed dissipative operators \cite{KrausZoller}.  Here, we
observe a trend towards a pure highly-entangled \textit{non}-equilibrium state
upon increasing the particle number $N$. An intriguing question arises as to
which values both characteristics then saturate in the thermodynamic limit $N
\rightarrow \infty$. Will, for example, the purity value approach unity? If
``yes'' then we would have a first example of dissipative engineering of a
time-periodic quantum state.  Unfortunately, it was not possible to go beyond
$N \approx 300$ by using the numerical spectral methods.

Furthermore, spectral properties of a Floquet map may also provide insight into
the relaxation towards the corresponding quantum attractor. A typical spectrum
of a map is shown in the inset of figure \ref{fig:bifurcation}(b). It has a
shape inherent to the spectra of completely positive trace-preserving maps
\cite{wolf1}. Namely, it has the spectral radius $1$, includes the single
eigenvalue $\lambda_1 = 1$, and is invariant under complex conjugation. The
spectral gap
\begin{equation}
\label{eq:gap}
\Delta = 1-|\lambda_2|,
\end{equation}
where $\lambda_2$ is the second largest eigenvalue by absolute value, can be
used to estimate the inverse relaxation time from a randomly chosen initial
state \cite{zyk,znidar,fazio0}. The spectral gap also exhibits a strong
dependence on the interaction strength; see figure \ref{fig:bifurcation}(b).

\begin{figure}
\begin{center}
\includegraphics[width=\textwidth]{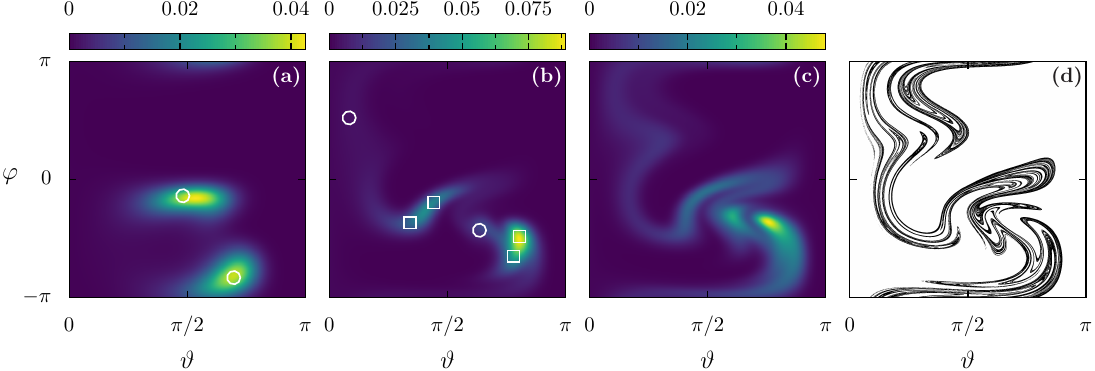}
\end{center}
\caption{(a-c) Poincar\'e-Husimi representation of the asymptotic operator
$\rho_a(0)$ for (a) $UN/J=0.2$, (b) $UN/J=0.8$ and (c) $UN/J=1$. Symbols
indicate attractors of the mean-field system, (\ref{eq:thetaphidot}),
period-two ($\ocircle$) and period-four ($\Box$) limit cycles. (d) Poincar\'e
section of the mean-field attractor for $UN/J=1$. The other parameters are
$\mu_0/J=1$, $\mu_1/J=3.4$, $\omega/J=1$, $\gamma N/J=0.1$, and $N=250$.}
\label{fig:poincare}
\end{figure}

\begin{figure}
\begin{center}
\includegraphics[width=0.99\textwidth]{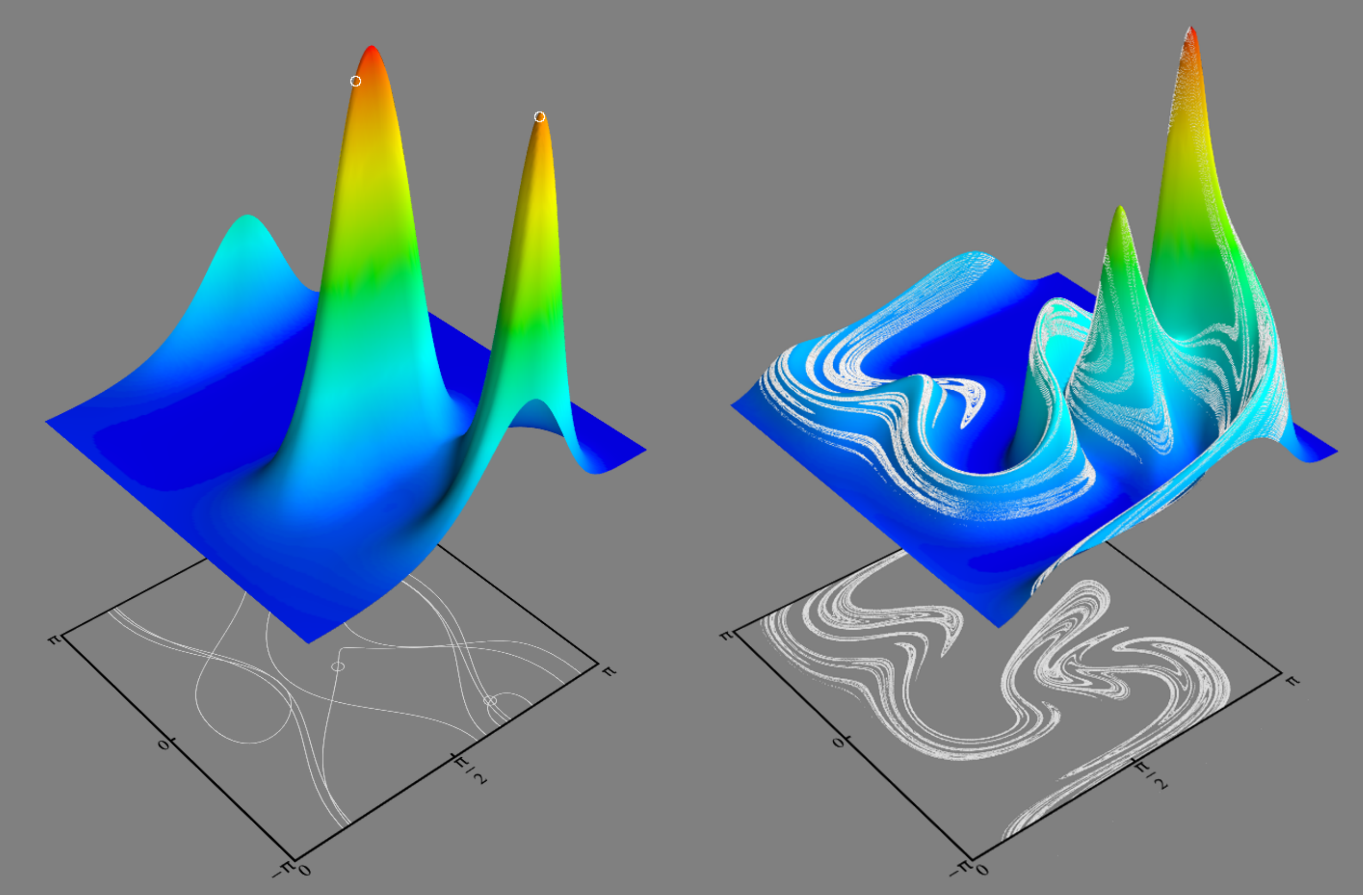}
\end{center}
\caption{3D versions of the Poincar\'e-Husimi representation of the asymptotic
states. Left panel corresponds to figure \ref{fig:poincare}(a) while right
panel corresponds to figure \ref{fig:poincare}(c,d). Bottom planes present the
Poincar\'e sections (dots) of the corresponding classical attractors (the line
on the left plane shows full period-two cycle).} \label{fig:3D}
\end{figure}

Different mean-field regimes can be visualized by plotting stroboscopic
Poincar\'e sections on the plane $\{\vartheta, \varphi\}$. In figure
\ref{fig:poincare}, classical Poincar\'e sections are compared with the
Poincar\'e-Husimi distributions $p(\vartheta,\varphi)$ of the quantum
asymptotic state obtained by projecting the density operator
$\varrho_{\mathrm{a}}(0)$ on the set of the generalized SU(2) coherent states
\cite{coherent1}. For $UN/J=0.2$ ($0.8$), the mean-field model predicts two
(six) points on the Poincar\'e section, corresponding to period-two (period-two
plus period-four) attractor(s); see symbols in figures \ref{fig:poincare}(a,b).
The Poincar\'e-Husimi distributions, figures \ref{fig:poincare}(a-b), reveal a
concentration of $p(\vartheta,\varphi)$ near these points. We attribute the
minor mismatch to finite-size effects. For $UN/J = 1$, the mean-field system
(\ref{eq:thetaphidot}) exhibits a chaotic attractor, figure
\ref{fig:poincare}(d), and the Poincar\'e-Husimi distribution, figure
\ref{fig:poincare}(c), fits the structure of this classical attractor for $N =
250$. Figure \ref{fig:3D} shows three-dimensional plots of the quantum
attractors super-imposed on the classical Poincar\'e sections both for the case
in which the mean-field equations predict two points (from a period-two limit
cycle) or a chaotic attractor, corresponding respectively to figures
\ref{fig:poincare}(a) and \ref{fig:poincare}(c,d).

It is noteworthy that the inverse particle number $1/N$ can be thought of as an
effective Planck constant, thus allowing for the comparison with the results
obtained for single-particle models \cite{Dittrich1,acta_pol, Dittrich2,
attractor1}; note in addition those cited in the mini-review~\cite{Dittrich2}.

\section{Existence of an effective time-independent generator}

\begin{figure}
\begin{center}
\includegraphics[width=0.6\textwidth]{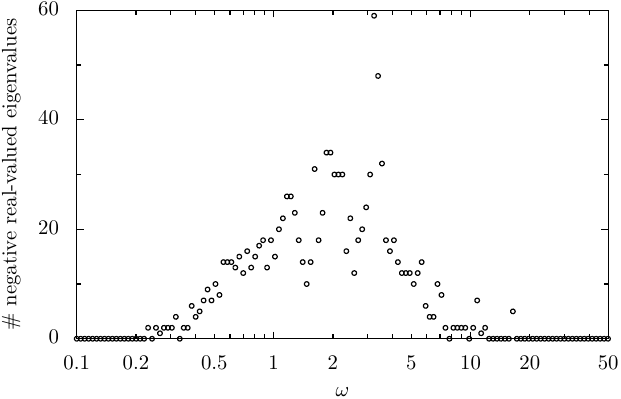}
\end{center}
\caption{Number of negative real-valued eigenvalues in the
spectrum of the Floquet map $\calP_{\mathrm{F}}$ as function of the driving
frequency $\omega$ (see text for definition). Whenever there is at least one
negative eigenvalue, the map does not allow for an effective time-independent
stroboscopic Lindblad generator $\calL_\eff$, such that $\calP_{\mathrm{F}} =
\exp(\calL_{\eff}T)$.  The parameters are $\mu_0/J=1$, $\mu_1/J=3.4$, $UN/J=1$,
$\gamma N/J=0.1$, and $N=30$.} \label{fig:4}
\end{figure}

The Floquet map $\calP_{\mathrm{F}}$ is a completely-positive and
trace-preserving map which belongs, following the nomenclature introduced in
\cite{inf}, to the class of time-dependent Markovian channels. It is an
interesting question whether it is possible to find an effective
time-independent generator $\calL_\eff$ \textit{of Lindblad form}
(\ref{lind}--\ref{dissipator}) that can mimic the action of the original
generator at stroboscopic instants of time, such that $\calP_{\mathrm{F}} =
\exp(\calL_{\eff}T)$. There are three necessary (and altogether sufficient)
conditions which any Lindblad generator has to fulfill: (i) trace preservation,
$\mathrm{tr} [\calL_{\eff}A] = \mathrm{tr[A]}$, (ii) Hermiticity preservation,
$(\calL_{\eff}A)^{\dagger} = \calL_{\eff}A$, if $A^{\dagger} = A$, and
(iii) so-called ``conditionally completely positiveness'' \cite{wolf2}.

This is in distinct contrast with the case of a unitary
evolution \cite{shirley,bukov}. The effective time-independent Hermitian
operator $H_{\eff}$ can always be obtained as the logarithm of the unitary
Floquet propagator $U_T = {\cal T} \exp\left[-\frac{\rmi}{\hbar} \int_0^T
H(\tau) \mathrm{d}\tau\right]$.  Moreover, not only its prinicpal branch but
\textit{any} branch of the logarithm yields a valid $H_\eff$. The non-unitary case is much more
restricted: Only the branch of the logarithm of a dissipative Floquet map which
produces an operator possessing properties (i--iii) yields a legitimate
Lindblad generator $\calL_\eff$ \cite{wolf2}.

Condition (i) holds by default if we start with a trace preserving map (which
is our case).  Condition (iii) can formally be checked
with the algorithm given in \cite{wolf2}. However, it is hardly realizable in practice when $M >
2$ because it involves repeated solution of an ${\cal O}(M^4)$ optimization
problem within the mixed-integer semidefinite programming framework
\cite{semi}. Condition (ii) is much easier to check. 

Hermiticity preservation demands that the spectrum $\{\lambda_j\}$ of the map
$\calP_{\mathrm{F}}$ is invariant under complex conjugation; in other words, it
should consist of non-negative real eigenvalues or/and of complex conjugated
pairs of eigenvalues. If there are \textit{negative} real-valued eigenvalues
(strictly speaking, of odd algebraic multiplicity) it is impossible to fulfill
the condition of the invariance of the spectrum under complex conjugation.
This is because any branch of the logarithm of a negative real-valued number
can neither produce a real number nor a complex conjugated pair.  Figure
\ref{fig:4} depicts the number of eigenvalues $\{\lambda_j\}$ with
$\mathrm{Re}(\lambda_j) < -\varepsilon$ and $|\mathrm{Im}(\lambda_j)| <
\varepsilon$, $\varepsilon = 10^{-7}$, as a function of the driving frequency
$\omega$. The dependence reveals that the condition is not fulfilled in the
most interesting case of non-\textit{adiabatic} and non-\textit{diabatic}
driving, when the asymptotic state of the dimer is sculpted by the modulations.
Apparently, an effective stroboscopic time-independent Lindblad generator does
not exist in this parameter region.

\section{Conclusions}

We demonstrated that the concept of dissipative Floquet maps provides an
operational way to identify `quantum attractors', i.e., asymptotic
time-periodic states of modulated open quantum systems, and estimate the
relaxation time towards them. To illustrate this idea, we have applied the
concept to a dissipative and periodically driven many-body model. We have
studied the model both exactly and, in the limit of a large particle number,
within a mean-field picture. The latter predicts bifurcations from regular to
chaotic attractors as the interaction strength is varied. The analysis shows a
strong dependence of quantum characteristics of the asymptotic non-equilibrium
many-body state, such as the purity and the negativity, on the interaction
strength, especially in proximity of bifurcations predicted by the mean-field
theory.

It is interesting to contrast the idea of Floquet maps produced by
time-periodic Lindblad generators, and an approximate Bloch-Redfield master
equation. The latter constitutes a well-known alternative to the Lindblad
formalism \cite{weiss}. Typically one then starts from a bilinear coupling of
the system to a heat bath of harmonic oscillators. The bath is characterized by
its spectral properties. When the model Hamiltonian is time-periodic, it is
possible, by assuming an Ohmic heat bath and following the Born-Markov
ideology, to derive the so-called Floquet-Markov equation \cite{grifoni,ketz1}.
This linear equation, similar to (\ref{lind}), governs the evolution of the
system density operator; it is also local in time with a time-periodic
generator. Formally it is thus possible to construct a corresponding Floquet
map in this case as well. The only problem is, however, this so obtained map
does not guarantee completely positivity for the evolution of the reduced
density operator \cite{v1}; even more problematic is that it may even not
necessarily assure the \textit{positivity} of the reduced density operator
towards its asymptotic limit, see \cite{v2,v3,juzar} for detailed comparisons.

We conclude by pointing out possible research directions which may benefit from
the use of Floquet maps within the Lindblad framework. It has been proposed to
use time-periodic driving to create, for the situation with coherent
Hamiltonian systems, effective topologically protected states
\cite{top,Kitagawa}. The important problems of the stability of these states
against dissipation or their creation with a synthetic dissiaption
\cite{fazio2} could be investigated by making use of our concept. Another
interesting question is whether the idea of `engineering by dissipation'
\cite{wolf,barr,kienz,kraus} can be extended to periodically modulated systems.
Finally, recent progress in the field of many-body localization (MBL)
inaugurates yet another potential application; the effect of temporal driving
on the localization has been addressed in \cite{ab1,laz,laz1} and, very
recently, the dynamics of open MBL systems was considered in
\cite{fish,les,les2}. We expect that the idea to combine the two latter
ingredients may soon invigorate the MBL community in pursuing future research
in this spirit; see also a very recent \cite{laz2}.

\ack 
The numerical simulations were supported by the Russian Science Foundation
Grant No.~15-12-20029 (M.\,I., S.\,D., and P.\,H.) and were performed on the
Lobachevsky cluster of the University of Nizhny Novgorod. D.\,P. acknowledge
the support by Singapore MOE Academic Research Fund Tier-2 project (Project
No.~MOE2014-T2-2-119, with WBS No.~R-144-000-350-112). S.\,D. and P.\,H.
acknowledge support by the Deutsche Forschungsgemeinschaft (DFG) via grants
(DFG) HA1517/35-1 (P.\, H.), DE1889/1-1 (S.\,D.). P.\,H. also acknowledges
support by the Singapore Ministry of Education and the National Research
Foundation of Singapore.

\section*{References}

\newcommand{\NatPhys}{{\it Nat. Phys.} }
\newcommand{\PRA}{{\it Phys. Rev. A} }
\newcommand{\PRB}{{\it Phys. Rev. B} }
\newcommand{\PRE}{{\it Phys. Rev. E} }
\newcommand{\Science}{{\it Science} }
\newcommand{\Nature}{{\it Nature} }
\newcommand{\PhysRep}{{\it Phys. Rep.} }
\newcommand{\AdvPhys}{{\it Adv. Phys.} }

\end{document}